\documentclass[a4paper,11pt,dvips]{article}
\usepackage{graphicx}
\usepackage{amsmath}
\usepackage{amssymb}
\usepackage{latexsym}
\usepackage[colorlinks]{hyperref}
\usepackage{color}
\usepackage{cite}
\usepackage{float}
\usepackage[textfont={footnotesize,sf},labelfont={color=blue,bf,sf},labelsep=endash]{caption}
\usepackage[position=top,labelfont={color=blue,bf,sf}]{subfig}
\usepackage{bm}
\usepackage{makeidx}
\usepackage{colortbl}

\newcommand{\bra}{\begin{array}}
\newcommand{\era}{\end{array}}
\newcommand{\beq}{\begin{equation}}
\newcommand{\eeq}{\end{equation}}
\newcommand{\beqar}{\begin{eqnarray}}
\newcommand{\eeqar}{\end{eqnarray}}

\def\BC{\bb C}
\def\_\BC{\bbi C}

\def\Tr {{\rm Tr}}

\def\( {\left(}
   \def\) {\right)}
\def\[ {\left[}
\def\] {\right]}

\def\Tr {{\rm Tr}}
\def\dag {{\dagger}}

\newcommand{\om}{\omega}

\newcommand{\lb}{\label}

\begin{document}
\begin{titlepage}
\setcounter{page}{1}
\renewcommand{\thefootnote}{\fnsymbol{footnote}}


\vspace{5mm}
\begin{center}

{\Large \bf {Thermodynamic Properties of Graphene in  Magnetic Field \\
and Rashba Coupling }}

\vspace{5mm}

{\bf Rachid Hou\c{c}a$^{a,b}$, Ahmed Jellal\footnote{a.jellal@ucd.ac.ma}$^{b}$}

\vspace{5mm}

{$^{a}$\em Equipe de Physique Th\'eorique et Hautes Energies,
Facult\'e des Sciences, Université Ibn Zohr},\\
{\em PO Box 8106, Agadir, Maroc}

{$^{b}$\em Laboratory of Theoretical Physics,  
Faculty of Sciences, Choua\"ib Doukkali University},\\
{\em PO Box 20, 24000 El Jadida, Morocco}

\vspace{3cm}

\begin{abstract}
We study the thermodynamic properties of massless Dirac fermions in graphene 
subjected to
a uniform magnetic field $B$ together with Rashba coupling parameter $\lambda_R$. 
The thermodynamic functions such as  the Helmholtz free energy,
total energy, entropy and heat capacity are obtained in the high temperature regime  using
an approach based on the zeta function. These  functions will be numerically examined  
by considering 
two cases related to $\lambda_R$ smaller or greater 
than $B$.
In particular,
we show that the  Dulong-Petit law is verified for  both cases.
\end{abstract}

\end{center}

\vspace{5cm}


\end{titlepage}

\section{Introduction}
Graphene is a two-dimensional system formed by  a plane of carbon atoms arranged
in a honeycomb structure \cite{Geim,Berger}. 
Since it isolation in 2004, the rise of this material is growing because   its 
electronic and mechanical properties are interesting. In particular,
graphene is the material having the greatest electrical mobility at
ambient temperature \cite{Bolotin,Chen}. It is also very flexible,
extremely strong and is an excellent thermal conductor. These properties
give graphene incredible potential for many applications in the fields of
electronics, composite materials, energy storage. The transition from laboratory
to industry is primarily based on the possibility of producing graphene on a large scale and at a reasonable cost.

Nowadays, graphene-based systems have no particular utility in magnetic information
storage applications or in spintronics \cite{Dery,Datta}, mainly because of
the absence of magnetic moments in carbon and the negligible interaction between
the spin of electrons with the movement of electronic charges (spin-orbit coupling)
\cite{Guinea}. This situation changes radically when graphene is interfaced with
other materials  by creating a heterostructure, that will offer the possibility
of manipulating the spin of the electron for graphene via the spin-orbit coupling
which can be controlled by the application of an electric field in the transverse
direction of gas \cite{rash,Bychkov}. This property has generated considerable
research activity due to possible applications in the emerging field of spintronics.
An interesting direction of research is to combine the spin-orbit coupling of Rashba
with the thermodynamic properties of graphene \cite{Balandin,Rusanov}. On the theoretical
level, it is necessary to better understand the combined effects of an intense electric
field of spin-orbit coupling, confinement and disorder potentials \cite{Nitta,Heida,Papadakis},
which have a direct bearing on the charge transport and spin properties of
the electron that reflect on thermodynamic quantities.

Thermodynamic properties of graphene based on the
statistical physics formalism was the subject of different investigations.
Indeed,
the thermodynamic properties of graphene nanoribbons under zero and finite magnetic fields have
been investigated \cite {Wright}. 
The electronic states in the tight-binding
approximation were used to calculate the thermal and magnetic properties. In the case of a finite
magnetic field, the Harper equation was solved for the electronic state for various wavevectors. The
magnetic field and temperature dependence of the magnetic susceptibility over a wide field and
temperature range were presented.
Thermodynamic properties of a weakly
modulated graphene monolayer in a
magnetic field was analyzed in \cite{Tahir}
and found that
 the modulation-induced effects on the thermodynamic properties are
enhanced and less damped with temperature in graphene compared with conventional 2DEG
systems.

We study the thermodynamic properties of Dirac fermions in graphene subjected to
an uniform magnetic field with the Rashba coupling parameter. After getting the solutions of
the energy spectrum, we use the zeta function to explicitly determine
the partition function that will be used to 
calculate the Helmholtz free energy,  total energy,  entropy and  specific heat. These
  will be numerically analyzed by considering two cases of $\lambda_R$  
smaller/greater than $B$ under suitable conditions of the physical
parameters. 
We show that 
the so-called Dulong-Petit law can be recovered  as a limiting case from our results. 

The manuscript is organized as follows. In section $2$, we formulate our problem by
introducing the bosonic operators and choosing a convenient gauge to
diagonalize the corresponding  Hamiltonian. 
The partition function of the system will be determined  via an approach
based on zeta function and  the  thermodynamic functions that describe the thermal physics
of monolayer graphene with Rashba coupling will be calculated in section $3$. 
Section 4 will be devoted to
the numerical results and discussions as well as comparison with literature.
We conclude our results in the final section.

\section{Theoretical model}

To study  the thermodynamic
proprieties of graphene subjected to an uniform magnetic field
with Rashba coupling, we consider
 the  tight-binding model together with Rashba term.
 This is governed by the Hamiltonian
\beq\lb{tigh}
H = H_0 + H_R
\eeq
where the  first term is given by \cite{Neto}
\beq
H_0=-t\sum_{\langle i,j\rangle,\alpha}a_{i,\alpha}^\dag a_{j,\alpha}
\eeq
 $a_{i,\alpha}^\dag$ and $a_{j,\alpha}$ are the creation and annihilation operators
of fermions placed on the site $i$,  $\alpha$ labels the spin and $t \simeq 2,8 eV$  is
the intralayer hopping parameter between nearest-neighbor sites. The Rashba Hamiltonian
for nearest-neighbor hopping reads as \cite{Qiao}
\beq
H_R=it_R\sum_{\langle i,j\rangle}a_{i}^\dag \vec{z}\left(\vec{s}\times \vec{d}_{i,j}\right) a_{j}
\eeq
where $\vec{d}_{i,j}$ is the lattice vector pointing from site $j$ to site $i$, $\vec{s}$
is a vector whose elements are the Pauli matrices in the
spin space, the spin-orbit coupling $t_R$ is determined by the strength of the electric field,
$a_{i}^\dag =\left(a_{i,\uparrow}^\dag,a_{i,\downarrow}^\dag\right)$ are vectors in the spin space
\cite{Konschuh}. Expanding the Hamiltonian
(\ref{tigh}) in the vicinity of the valleys $\vec{K}_\pm = (\pm\frac{4\pi}{3}\sqrt{3}, 0)$, gives the
low-energy Hamiltonian in the sublattice
$(A,B)$ and spin $(\uparrow,\downarrow)$ spaces
\beq\lb{HH}
H=v_F\left(\eta\sigma_x  \pi_x+\sigma_y  \pi_y\right)+\lambda_R\left(\eta\sigma_xs_y -\sigma_ys_x s\right)
\eeq
where  the conjugate momentum $\pi_x$ and $\pi_y$ can be written in symmetric gauge $\vec{A}=\frac{B}{2}(-y,x)$ as
\beq
\pi_x=p_x-\frac{eB}{2}y,\qquad \pi_y=p_y+\frac{eB}{2}x.
\eeq
the Fermi velocity $v_F=\frac{3at}{2\hbar}$ with $a=1.142 nm$, $\eta = \pm1$ labels the valley degrees
of freedom, $\sigma=\left(\sigma_x,\sigma_y\right)$ are the Pauli matrices of pseudospin operator
on $A(B)$ lattice cites. Note that the present system also presents the intrinsic spin orbit coupling
(SOC), but its value is very weak  $\sim 12\mu eV$ \cite{Konschuh} compared to
the Rashba coupling parameter $\lambda_R=\frac{\hbar v eE}{4mc^2}$ \cite{kane}, which depends on the external
electric field perpendicular to the graphene plane is very strong compared to intrinsic SOC.

The Hamiltonian (\ref{HH}) around a single Dirac point $(\eta = +1)$ on the double-spinor basis
$|\Psi\rangle = ( |r_A,n,\uparrow\rangle,|r_B,n-1,\downarrow\rangle,|r_B,n,\uparrow\rangle,|r_A,n-1,\downarrow\rangle)^t$ 
is given by
 \beq\lb{hhi}
H=\left(
\begin{array}{cccc}
 0 & 0 & v\left(\pi_x-i\pi_y \right) & 0 \\
 0 & 0 & 0 & v\left(\pi_x+i\pi_y \right) \\
 v\left(\pi_x+i\pi_y \right) & 0 & 0 & -2i\lambda_R  \\
 0 & v\left(\pi_x-i\pi_y \right) & 2i\lambda_R  & 0 \\
\end{array}
\right).
\eeq
To diagonalize the Hamiltonian (\ref{hhi}), it is convenient to  introduce the usual bosonic operators
in terms of the conjugate momentum
\beq
a=\frac{\ell_B}{\sqrt{2}\hbar}\left(\pi_x-i\pi_y\right),\qquad a^\dag=\frac{\ell_B}{\sqrt{2}\hbar}\left(\pi_x+i\pi_y\right)
\eeq
which verify the commutation relation $[a,a^\dag]=\mathbb{I}$,  $\ell_B = \sqrt{\frac{\hbar}{eB}}$  is the magnetic length.
Mapping (\ref{hhi}) in terms of $a$ and $a^\dag$ to end up with
\beq
H=\left(
 \begin{matrix}
 0 & 0 & \frac{\sqrt{2} \hbar v}{\ell_B} a & 0 \\
 0 & 0 & 0 & \frac{\sqrt{2} \hbar v}{\ell_B} a^\dag \\
 \frac{\sqrt{2} \hbar v }{\ell_B}a^\dag & 0 & 0 & -2i\lambda_R  \\
 0 & \frac{\sqrt{2} \hbar v}{\ell_B} a & 2i\lambda_R  & 0 \\
\end{matrix}
\right).
\eeq
To obtain the solution of the energy spectrum
we  act the Hamiltonian on the state $|\Psi\rangle$  leading to the eigenvalue equation
\beq
\left(
 \begin{matrix}
 -E & 0 & \frac{\sqrt{2} \hbar v}{\ell_B} a & 0 \\
 0 & -E & 0 & \frac{\sqrt{2} \hbar v}{\ell_B} a^\dag \\
 \frac{\sqrt{2} \hbar v }{\ell_B}a^\dag & 0 & -E & -2i\lambda_R  \\
 0 & \frac{\sqrt{2} \hbar v}{\ell_B} a & 2i\lambda_R  & -E \\
\end{matrix}
\right)\left(
         \begin{matrix}
           |r_A,n,\uparrow\rangle \\
           |r_B,n-1,\downarrow\rangle \\
           |r_B,n,\uparrow\rangle \\
           |r_A,n-1,\downarrow\rangle \\
         \end{matrix}
       \right)=\left(
                 \begin{matrix}
                   0 \\
                   0 \\
                   0 \\
                   0 \\
                 \end{matrix}
               \right)
\eeq
giving rise to  the following set
\begin{eqnarray}
 && -E|r_A,n,\uparrow\rangle+\frac{\sqrt{2} \hbar v}{\ell_B}a|r_B,n,\uparrow\rangle = 0 \\
  &&-E|r_B,n-1,\downarrow\rangle+\frac{\sqrt{2} \hbar v}{\ell_B} a^\dag|r_A,n-1,\downarrow\rangle = 0 \\
  && \frac{\sqrt{2} \hbar v}{\ell_B} a^\dag|r_A,n,\uparrow\rangle-E|r_B,n,\uparrow\rangle-2i\lambda_R|r_A,n-1,
  \downarrow\rangle = 0
  \\
  && \frac{\sqrt{2} \hbar v}{\ell_B}a|r_B,n-1,\downarrow\rangle+2i\lambda_R|r_B,n,\uparrow\rangle)-E|r_A,n-1,\downarrow\rangle =0.
\end{eqnarray}
These can be solved to obtain a second order equation for the eigenvalues 
\beq
E^2\pm2\lambda_RE-\left(\hbar\omega_D\right)^2n=0,\qquad n=0,1,2\cdots
\eeq
where $\omega_D=v_F\sqrt{\frac{2 eB}{\hbar }}$ is the Dirac constant. A straightforward calculation leads to the following solutions
\begin{eqnarray}\lb{eee}
  && E_{1,n}^\pm = -\lambda_R\pm\sqrt{\lambda_R^2+\left(\hbar\omega_D\right)^2n} \\
  && E_{2,n}^\pm = \lambda_R\pm\sqrt{\lambda_R^2+\left(\hbar\omega_D\right)^2n}
\end{eqnarray}
These are four band solutions and are strongly dependent on the Rashba coupling parameter $\lambda_R$,
which 
reduces to two bands once $\lambda_R$ is switched off. As it will be shown, this parameter
will play a crucial role in the forthcoming study.
Next we will see how the above results can be employed 
to explicitly determine in the first stage the partition function and in the second one 
derive the related 
thermodynamic functions. 


\section{Thermodynamic functions}


We will study the thermodynamic properties of the monolayer graphene with Rashba coupling in contact
with a thermal reservoir at finite temperature. For  simplicity, we assume that only fermions with positive energy
$(E > 0)$ are regarded to constitute the thermodynamic ensemble \cite{Santos}. We start by evaluating
the corresponding partition function
\beq\lb{ZZe}
\mathbb{Z}=\Tr e^{-\beta H}=
\sum_{n=0}^{+\infty}\left(e^{-\beta E_{1,n}^+}+e^{-\beta E_{2,n}^+}\right)
\eeq
where $\beta=\frac{1}{k_BT}$, $k_B$ is the Boltzmann constant and $T$ is the equilibrium temperature.
Using (\ref{eee}-\ref{ZZe}) and after some algebra, we show that $\mathbb{Z}$ takes the form
\beq
\mathbb{Z}=2\cosh(\beta\lambda_R)\sum_{n=0}^{+\infty}e^{-\beta\sqrt{\left(\hbar\omega_D\right)^2n+\lambda_R^2}}.
\eeq
To go further, let us make the following changes 
\beq
r=\left(\frac{\hbar\omega_D}{\lambda_R}\right)^2,\qquad \gamma=\sqrt{r}=\frac{\hbar\omega_D}{\lambda_R},\qquad
\kappa=\frac{1}{r},\qquad \tau=\frac{k_BT}{\lambda_R}
\eeq
which can be used to write $\mathbb{Z}$  as
\beq\lb{z}
\mathbb{Z}=2\cosh(\beta\lambda_R)\sum_{n=0}^{+\infty}e^{-\frac{\gamma}{\tau}\sqrt{n+\kappa}}.
\eeq
Now considering 
the integral formula \cite{Ma1}
\beq
e^{-x}=\frac{1}{2\pi i}\int_Cx^{-s}\Gamma(s)ds
\eeq
to convert  the sum in (\ref{z}) as 
\beq\lb{sum}
\sum_{n=0}^{+\infty}e^{-\frac{\gamma}{\tau}\sqrt{n+\kappa}}=\frac{1}{2\pi i}\int_C\left(\frac{\gamma}{\tau}\right)^{-s}
\sum_n(n+\kappa)^{-\frac{s}{2}}\Gamma(s)=\frac{1}{2\pi i}
\int_C\left(\frac{\gamma}{\tau}\right)^{-s}\sum_n\xi_H\left(\frac{s}{2},\kappa\right)\Gamma(s)
\eeq
with two zeta functions of Euler $\Gamma$ and Hurwitz $ \xi_H$.
The function 
$\xi_H$ is defined for complex arguments $s$ (with $\Re(s) > 1$) and $\kappa$  (with $\Re(\kappa) > 0$),
it is absolutely convergent for  given values of $s$ and $\kappa$, which can be extended to a
meromorphic function defined for all $s\neq1$. To calculate the sum (\ref{sum})
we apply
the residue theorem for the  poles $s = 0$
and $s = 2$ 
to get 
\beq
\sum_{n=0}^{+\infty}e^{-\frac{\gamma}{\tau}\sqrt{n+\kappa}}=\left(\frac{\tau}{\gamma}\right)^2+\xi_H\left(0,\kappa\right)
\eeq
where $\xi_H\left(0,\kappa\right)=\frac{1}{2}-\kappa$. Combining all to end up with  the partition function
\beq
\mathbb{Z}=2\left(\frac{(k_BT)^2-\lambda_R^2}{(\hbar\omega_D)^2}+\frac{1}{2}\right) \cosh \beta\lambda_R.
\eeq

By virtue  of the stability of the graphene at high temperature \cite{Kwanpyo} we will study
its thermodynamic properties to make a comparison with other studies. Indeed,
in the limit  $k_BT\gg\lambda_R$, the total  partition function $\mathbb{Z}_N$
for $N$ fermions can be approximated by
\beq\lb{ZZ}
\mathbb{Z}_N\simeq\left(\left(2\left(\frac{1 }{\beta\hbar\omega_D }\right)^2+1\right)  \cosh \beta\lambda_R\right)^N.
\eeq
We point out that several studies  have been developed on graphene  using  formalism based on zeta function. Among
them, we  cite work reported by Beneventano {\it et al.} \cite{Bene1} in studying the quantum Hall
effect, Berry's phases and minimal conductivity of graphene.
Also Falomir {\it et al.} \cite{Falomir}
evaluated
the Hall conductivity of the model  from the partition function
employing the zeta function approach to the associated functional determinant.

Since we have derived the partition function of our system, we can now determine all related thermodynamic
functions.
Indeed, after some algebras we obtain the Helmholtz free energy
$F$, total energy $U$, entropy $S$ and  specific heat $C$ 
\begin{eqnarray}\lb{ter}
 && \frac{F}{N} = -\frac{1}{\beta}\ln \mathbb{Z}=-\frac{1}{\beta }\ln\left(\frac{2}{\beta ^2 \hbar^2\omega_D ^2}+1\right)
  -\frac{1}{\beta }\ln\left(\cosh (\beta  \lambda_R )\right) \\ \
 && \frac{U}{N} = -\frac{\partial}{\partial \beta}\ln \mathbb{Z}=\frac{4-\beta
  \lambda_R  \left(\beta ^2 \hbar\omega_D ^2+2\right) \tanh (\beta  \lambda_R )}{\beta
  \left(\beta ^2 \hbar^2\omega_D ^2+2\right)} \\
   && \frac{S}{N k_B} =  \beta^2\frac{\partial F}{\partial \beta}=-\beta
  \lambda_R\left(\tanh (\beta  \lambda_R )\right)+\ln\left(\frac{2}{\beta ^2 \hbar^2\omega_D ^2}+1\right)
  \cosh (\beta  \lambda_R )+\frac{4}{\beta ^2 \hbar^2\omega_D ^2+2}\\
  && \frac{C}{Nk_B} = -\beta^2\frac{\partial U}{\partial \beta}=\frac{\beta ^2 \lambda_R ^2
  \left(\beta ^2 \hbar^2\omega_D ^2+2\right)^2 \text{sech}^2(\beta  \lambda_R )+12
  \beta ^2 \hbar^2\omega_D ^2+8}{\left(\beta ^2 \hbar^2\omega_D ^2+2\right)^2}.
\end{eqnarray}
Next, we will numerically analyse the above thermodynamic functions to underline the behavior
of our system. This will be done by giving some plots under suitable conditions
and making different discussions. 

\section{Numerical results and discussions}

\begin{figure}[!ht]
  \centering
  \includegraphics
  [width=16cm,  height=12cm]{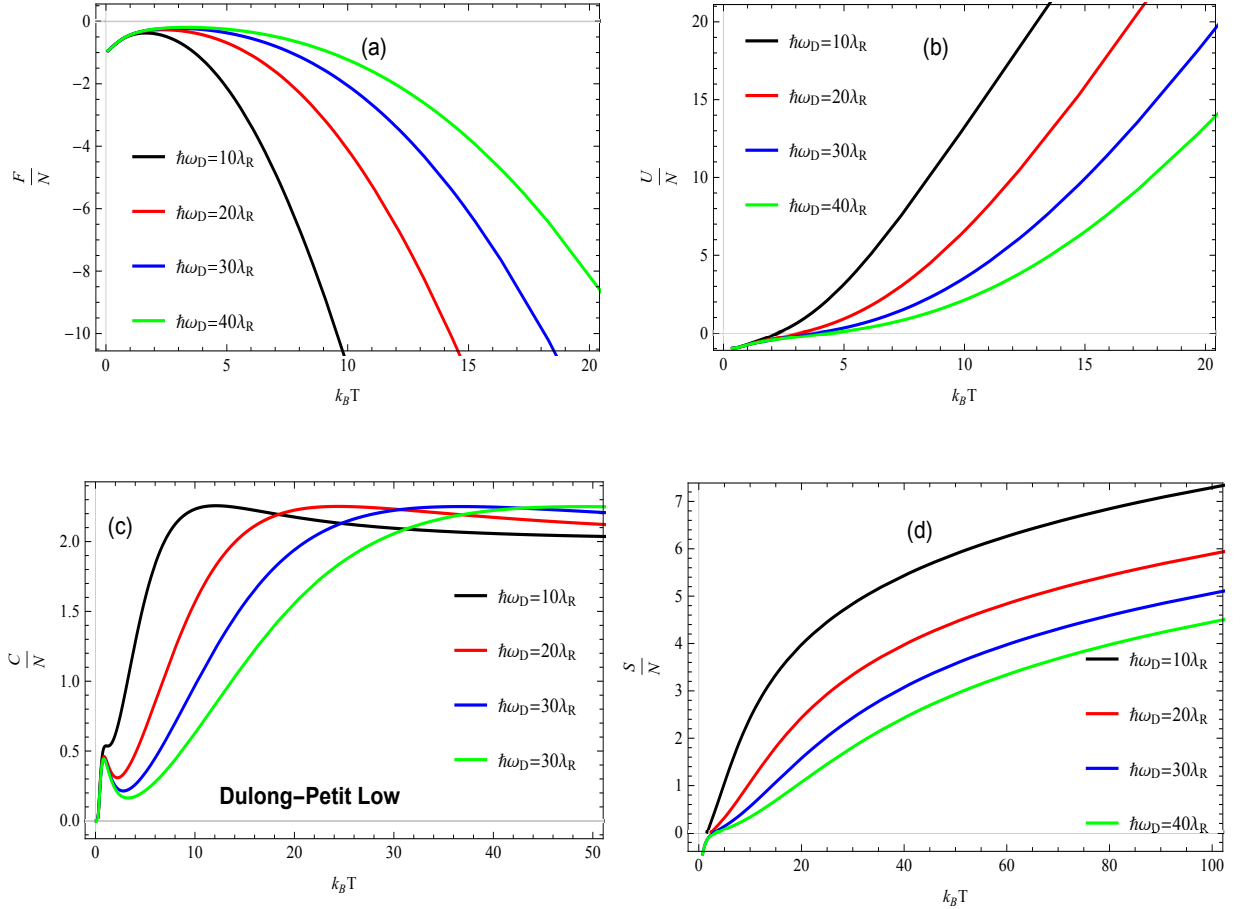}
  \caption{ (Color online) Thermodynamic functions of
  Dirac fermions in graphene with Rashba coupling $\lambda_R$ and magnetic field $B$
  versus the temperature $k_B T$
  for the values
  $\hbar\omega_D=10\lambda_R,20\lambda_R,30\lambda_R,40\lambda_R$. (a): The Helmholtz free energy, (b): the mean energy,
  (c): the entropy, (d): the heat capacity.}\label{fig}
\end{figure}

We recall our theory includes two interesting quantities, which
are the Rashba coupling parameter $\lambda_R$ and the external magnetic field $B$.
To carry out our numerical study, we consider the case when 
the ratio $\frac{\lambda_R}{\hbar \om_D}$ is smaller or  greater than one.
Indeed,
%
%
For a smaller ratio than one, 
%
%
the thermodynamic functions versus 
the temperature $k_BT$ for the fixed values $\hbar\omega_D=10\lambda_R, 20\lambda_R,30\lambda_R, 40\lambda_R$.  
It is clearly seen that in Figure \ref{fig}(a)  the Helmontz function $\frac{F}{N}$ has a little increase when
$k_BT$ starts to increase,
but it decreases rapidly for high values of $k_BT$. For the four fixed values $\frac{F}{N}$
is showing the same behavior except that the velocity of decrease changes, which means
that we can increase/decrease $\frac{F}{N}$ by tuning on $\lambda_R$.
%
In Figure \ref{fig}(b) for the interval $0< k_BT <5$ we observe that 
the behavior of the total energy $\frac{U}{N}$ 
is independent on 
different
 values of $\lambda_R$ while in the interval $5< k_BT <40$ it starts increasing with
a nearly linear behavior. The entropy $\frac{S}{N}$  is slowly increasing
for large $k_BT$ and decreases for strong values of  $\lambda_R$ as shown in Figure \ref{fig}(c).
In Figure \ref{fig}(d) the heat capacity $\frac{C}{Nk_B}$  is showing that there is a peak appearing
with small values of $\lambda_R$ and it tends to an asymptotic behavior fixed in the value $2$ when $k_BT$ increases.

\begin{figure}[!ht]
  \centering
  \includegraphics[width=16cm,  height=12cm]{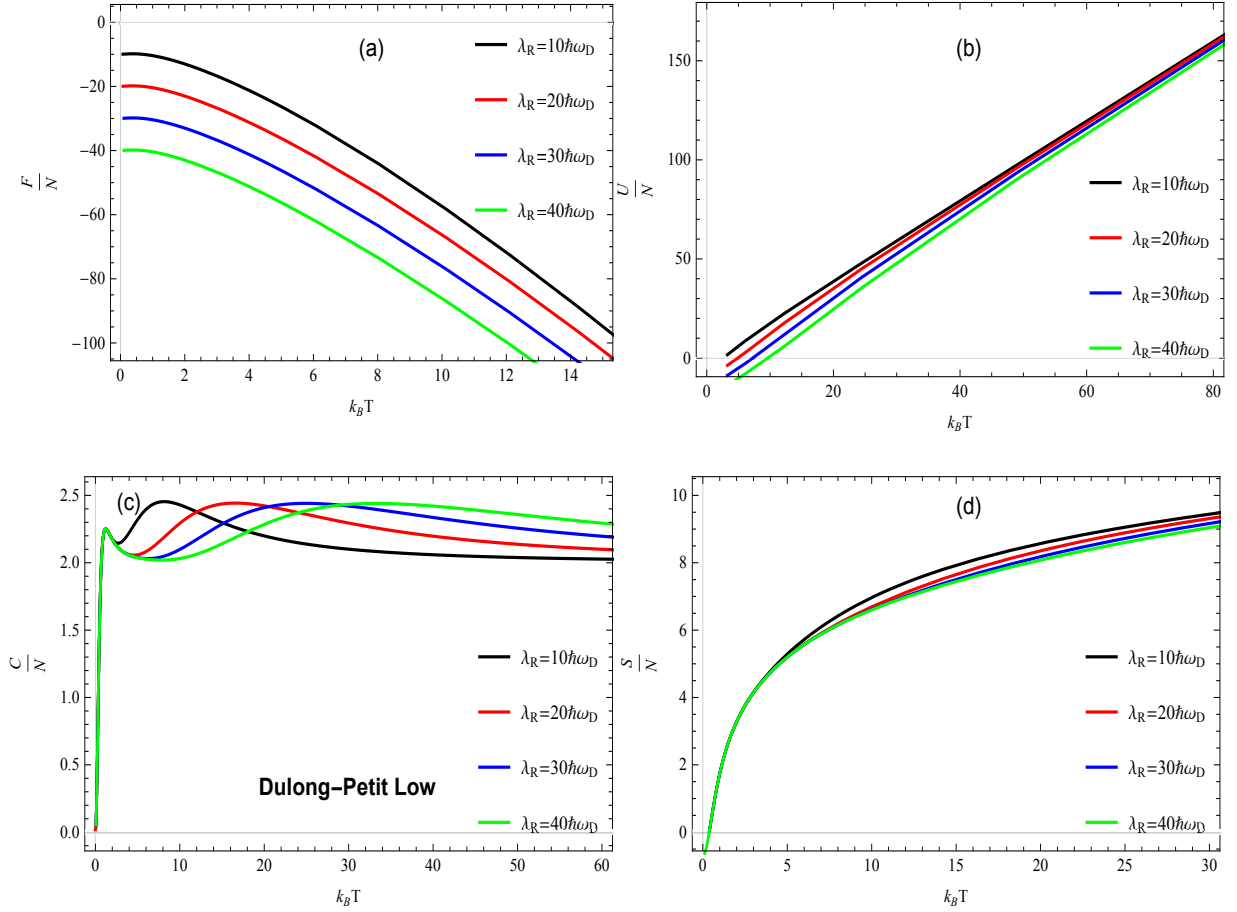}\\
  \caption{(Color online) Thermodynamic functions of
  Dirac fermions in graphene with Rashba coupling $\lambda_R$ and magnetic field $B$
  versus the temperature $k_B T$
  for the values
  $\lambda_R=10\hbar\omega_D,20\hbar\omega_D,30\hbar\omega_D,40\hbar\omega_D$. (a): The Helmholtz free energy, (b): the mean energy,
  (c): the entropy, (d): the heat capacity.}\label{fig2}
\end{figure}

In Figure \ref{fig2} we investigate the second case by shooing the values of the Rashba coupling parameters
$\lambda_R=10\hbar\omega_D$, $20\hbar\omega_D$, $30\hbar\omega_D, 40\hbar\omega_D$.
From Figure \ref{fig2}(a) We observe that the Helmontz function $\frac{F}{N}$  decreases when  
$k_BT>2$
and changes as long as $\lambda_R$ increases. 
The total energy
$\frac{U}{N}$ increases with  almost a linear  behavior when $0<k_BT<20$ but faraway 
we observe   that it shows
the same behavior for  any value of $\lambda_R$ as depicted in Figure \ref{fig2}(b). 
In Figure \ref{fig2}(c), the entropy $\frac{S}{N}$ is independent on the parameter $\lambda_R$
in the interval $0<k_BT<5$,
but when $k_BT>5$,  $\frac{S}{N}$  shows a small variation in terms of
$\lambda_R$. The heat capacity $\frac{C}{Nk_B}$ does not change with respect to the first case,
which has constantly  asymptotic behavior in the value $2$ for the high temperature regime
as shown in Figure \ref{fig2}(d).\\


By examining the thermodynamic properties for monolayer graphene with Rashba coupling,
we observe that the average energy for our system presents a critical point for low temperature,
that is, near $T=0K$. However,  the temperature of the system increases, the amount of free energy
available to perform the ensemble work decreases rapidly. In contrast, we notice that when this ensemble
tends to the thermal equilibrium in the reservoir, the internal energy of the system increases continuously.
The thermal variation of the system  tends to increase until reaching the thermal equilibrium, where it
becomes constant. Moreover, based on the solid state physics, we notice that the well-known Dulong-Petit
law is satisfied for our system, which is characterized by $\frac{C}{N}= 2k_B$ \cite{Santos} 
as seen in both Figures \ref{fig}(c) and
\ref{fig2}(c).

\section{Conclusion}

We have studied the thermodynamic properties of  a massless Dirac fermions in graphene
subjected to an uniform magnetic field
with Rashba interaction. 
The annihilation and creation operators were introduced to 
obtain the solutions of
the energy spectrum.
These are used together with a method based on the zeta function
to explicitly determine the partition function. Therefore
the thermodynamic functions such as
the Helmholtz free energy, total energy,
   entropy and heat capacity were obtained  in terms of the Rashba
coupling parameter.

Subsequently, two cases have been studied related to Rashba coupling parameter $\lambda_R$ 
and magnetic filed $B$ in the high
temperature regime. Indeed, we have analyzed numerically the case when $\lambda_R$ 
is smaller than  $B$, which  allowed us to recover a result already obtained  
in \cite{Santos} for graphene system. We have also considered the case when $\lambda_R$ 
is greater than $B$ where it was shown that the Dulong– Petit law  is always verified by our system.


\end{document}